\documentclass[10pt]{iopart}
\usepackage{graphicx, epsfig, epstopdf, float, subfig, multirow, amssymb}
\usepackage[draft]{hyperref}
\usepackage{grffile, pdflscape, rotating}

\newcommand{\pset}{\vartheta}

\newcommand{\likelihood}{p(\vec{d}(t)|\vec{\pset},H)}
\newcommand{\prior}{p(\vec{\pset}|H)}
\newcommand{\evidence}{p(\vec{d}(t)|H)}
\newcommand{\posterior}{p(\vec{\pset}|\vec{d}(t),H)}

\usepackage{color}
\usepackage{fancyvrb}

\usepackage[labelformat=parens,labelsep=quad,skip=3pt]{caption}

\begin{document}

\title{PyCBC Inference: A Python-based parameter estimation toolkit for compact binary coalescence signals}

\author{C.~M.~Biwer}\address{Los Alamos National Laboratory, Los Alamos, NM 87545, USA}\address{Department of Physics, Syracuse University, Syracuse, NY 13244, USA}
\author{Collin~D.~Capano}\address{Albert-Einstein-Institut, Max-Planck-Institut f\"ur Gravitationsphysik, D-30167 Hannover, Germany}
\author{Soumi~De}\address{Department of Physics, Syracuse University, Syracuse, NY 13244, USA}
\author{Miriam~Cabero}\address{Albert-Einstein-Institut, Max-Planck-Institut f\"ur Gravitationsphysik, D-30167 Hannover, Germany}
\author{Duncan~A.~Brown}\address{Department of Physics, Syracuse University, Syracuse, NY 13244, USA}
\author{Alexander~H.~Nitz}\address{Albert-Einstein-Institut, Max-Planck-Institut f\"ur Gravitationsphysik, D-30167 Hannover, Germany}
\author{V.~Raymond}\address{Albert-Einstein-Institut, Max-Planck-Institut f\"ur Gravitationsphysik, D-14476 Potsdam, Germany}\address{School of Physics and Astronomy, Cardiff University, Cardiff, CF243AA, Wales, UK}

\begin{abstract}
We introduce new modules in the open-source PyCBC gravitational-wave astronomy toolkit that implement Bayesian inference for compact-object binary mergers. We review the Bayesian inference methods implemented and describe the structure of the modules. 
We demonstrate that the PyCBC Inference modules produce unbiased estimates of the parameters of a simulated population of binary black hole mergers. We show that the posterior parameter distributions obtained used our new code agree well with the published estimates for binary  black  holes in the  first LIGO-Virgo observing run.
\end{abstract}

\maketitle
\ioptwocol

\section{Introduction}

The observations of six binary black hole
mergers~\cite{TheLIGOScientific:2016pea,Abbott:2017vtc,Abbott:2017gyy,Abbott:2017oio},
and the binary neutron star merger GW170817~\cite{TheLIGOScientific:2017qsa}
by Advanced LIGO~\cite{TheLIGOScientific:2014jea} and
Virgo~\cite{TheVirgo:2014hva} have established the field of gravitational-wave
astronomy. Understanding the origin, evolution, and physics of
gravitational-wave sources requires accurately measuring the properties of
detected events. In practice, this is performed using Bayesian
inference~\cite{Bayes:1763,Jaynes:2003jaq}. Bayesian inference allows us to determine the
signal model that is best supported by observations and to obtain posterior
probability densities for a model's parameters, hence inferring the
properties of the source.  In this paper, we present \emph{PyCBC Inference}; a
set of Python modules that implement Bayesian inference in the PyCBC
open-source toolkit for gravitational-wave
astronomy~\cite{alex_nitz_2018_1208115}.  PyCBC Inference has been used to
perform Bayesian inference for several 
astrophysical problems, including: testing the black hole area increase
law~\cite{PhysRevD.97.124069}; combining multi-messenger
obervations of GW170817 to constrain the viewing angle of the
binary~\cite{Finstad:2018wid}; and determining that the gravitational-wave
observations of GW170817 favor a model where both compact objects have the
same equation  of state, and measuring the tidal deformabilities and radii of
the neutron stars~\cite{De:2018uhw}. 

We provide a comprehensive
description  of the methods and code implemented in PyCBC Inference. We then
demonstrate that PyCBC Inference can produce unbiased estimates of the
parameters of a simulated population of binary black holes. We show that PyCBC
Inference
can recover posterior probability  distributions that are in good agreement
with  the  published measurements  of the binary black holes detected  in the
first LIGO-Virgo observing run~\cite{TheLIGOScientific:2016pea}.
This paper is organized as follows: Sec.~\ref{sec:bayes} gives an overview of
the Bayesian inference methods used in gravitational-wave astronomy for
compact-object binary mergers. We provide an overview of the waveform models
used; the likelihood function for a known signal in stationary, Gaussian
noise; the sampling methods used to estimate the posterior probability
densities and the evidence; the selection of independent samples; and the
estimation of parameter values from posterior probabilities.
Sec.~\ref{sec:code} describes the design of the PyCBC Inference software and
how the methods described in  Sec.~\ref{sec:bayes} are implemented in the
code.  Sec.~\ref{sec:validation} uses a simulated  population of binary black
holes and the black-hole mergers  detected in the first LIGO-Virgo observing
run to demonstrate the use of PyCBC Inference.  We provide the posterior
probability densities for the events GW150914, GW151226,  and
LVT151012, and the command lines and configurations to reproduce these results
as supplemental materials \cite{ResultsRepo}.  Finally, we summarize the
status of the code and possible future developments in Sec.~\ref{sec:con}.

\section{Bayesian Inference for Binary Mergers}\label{sec:bayes}

In gravitational-wave astronomy, Bayesian methods are used to infer the
properties of detected astrophysical
sources~\cite{Finn:1992xs,Cutler:1994ys,Nicholson:1997qh,Christensen:2001cr}.
Given the observed data $\vec{d}(t)$---here this is data from a
gravitational-wave detector network in which a search has
identified a signal~\cite{Allen:2005fk,Usman:2015kfa,Nitz:2017svb}---Bayes'
theorem~\cite{Bayes:1763,Jaynes:2003jaq} states that for a hypothesis $H$,
\begin{equation}
\label{eqn:bayes} \posterior =
\frac{\likelihood \prior}{\evidence} .
\end{equation} 
In our case, hypothesis $H$ is the model of the gravitational-wave signal 
and $\vec{\pset}$ are the parameters of
this model. Together, these describe the properties of the astrophysical
source of the gravitational waves.
In Eq.~(\ref{eqn:bayes}), the prior
probability density $\prior$ describes our knowledge about the
parameters before considering the observed data $\vec{d}(t)$, and the
likelihood $\likelihood$ is the probability of obtaining the observation
$\vec{d}(t)$ given the waveform model $H$ with parameters $\vec{\pset}$.

Often we are only interested in a subset of the parameters $\vec{\pset}$.
To obtain a probability distribution on one or a few parameters, we marginalize
the posterior probability by integrating $\likelihood \prior$ over the unwanted
parameters. Marginalizing over all parameters yields the evidence, $\evidence$,
which is the denominator in Eq.~(\ref{eqn:bayes}). The
evidence serves as a normalization constant of the posterior probability for
the given model $H$. If we have two competing models $H_{A}$ and $H_{B}$, the
evidence can be used to determine which model is favored by the data via the Bayes
factor~\cite{KassRaftery:1995,GelmanMeng:1998,skilling2006},
\begin{equation}
\mathcal{B} = \frac{p(\vec{d}(t)|H_{A})}{p(\vec{d}(t)|H_{B})}  .
\end{equation}
If $\mathcal{B}$ is greater than 1 then model $H_A$ is
favored over $H_B$, with the magnitude of $\mathcal{B}$ indicating 
the degree of belief.

PyCBC Inference can compute Bayes factors and produce marginalized
posterior probability densities given the data
from a network of gravitational-wave observatories with $N$ detectors
$\vec{d}(t) = \{d_{i}(t); 1 < i < N\}$, and a model $H$ that describes the
astrophysical source.  In the remainder of this section, we review the methods
used to compute these quantities.

\subsection{Waveform Models}\label{sec:models}

The gravitational waves radiated in  a binary merger's source frame  are described by
the component masses $m_{1,2}$, the three-dimensional spin vectors
$\vec{s}_{1,2}$ of the compact objects~\cite{Hawking:1987en}, and the
binary's eccentricity $e$~\cite{Peters:1964zz}.  A parameter $\phi$ describes
the phase of the binary at a fiducial reference time, although this is not
usually of physical interest. For binaries containing neutron stars,
additional parameters $\Lambda_{1,2}$ describe the star's tidal
deformabilities~\cite{Flanagan:2007ix,Hinderer:2007mb}, which depend on the
equation of state of the neutron stars.  The waveform observed by the
Earth-based detector network depends on seven additional parameters: the
signal's time of arrival $t_c$, the binary's luminosity distance
$d_\mathrm{L}$, and four Euler angles that describe the transformation from the
binary's frame to the detector network frame~\cite{Wahlquist:1987rx}.  These
angles are typically written as the binary's right ascension $\alpha$,
declination $\delta$, a polarization angle $\Psi$, and the inclination angle
$\iota$ (the angle between the binary's angular momentum axis and the line of
sight).  

Binary mergers present a challenging problem for Bayesian inference, as the
dimensionality of the signal parameter space is large. This is further
complicated by correlations between the signal's parameters. For example, at
leading order the gravitational waveform depends on the chirp mass
$\mathcal{M}$~\cite{PhysRev.131.435}. The mass ratio enters the waveform at
higher orders and is more difficult to measure. This results in an
amplitude-dependent degeneracy between the component
masses~\cite{Christensen:2001cr}. Similarly, the binary's mass ratio can be
degenerate with its spin~\cite{Hannam:2013uu}, although this degeneracy can be
broken if the binary is precessing. Much of the effort of parameter estimation
in gravitational-wave astronomy has focused on developing computationally
feasible ways to explore this signal space, and on extracting physically
interesting parameters (or combinations of parameters) from the large,
degenerate parameter space (see e.g. Ref.~\cite{Veitch:2014wba} and references therein).  However, in many problems of interest, we are not
concerned with the full parameter space described above.  For example, field
binaries are expected to have negligible eccentricity when they are observed
by LIGO and Virgo~\cite{PhysRev.131.435}, and so eccentricity  is neglected in
the waveform models. Simplifying assumptions can be made about the compact
object's spins (e.g. the spins are aligned with the binary's orbital angular
momentum), reducing the dimensionality of the waveform parameters
space.

Given a set of parameters $\vec{\pset}$, one can obtain a model of the
gravitational-wave signal from a binary merger using a variety of different
methods, including: post-Newtonian theory (see e.g. Ref.~\cite{Blanchet2006}
and references therein), analytic models calibrated against numerical
simulations~\cite{Buonanno:1998gg,Buonanno:2000ef,Damour:2000we,Damour:2001tu,Ajith:2007qp,Ajith:2009bn,Santamaria:2010yb},
perturbation theory~\cite{Teukolsky:1972my,Berti:2009kk}, and full numerical
solution of the Einstein equations (see e.g. Ref.~\cite{Cardoso:2014uka} and
references therein). Obtaining posterior probabilities and evidences can
require calculating $\mathcal{O}\left(10^9\right)$ template waveforms,
which restricts us to models that are computationally efficient to calculate.
The cost of full numerical simulations makes them prohibitively expensive at
present. Even some analytic models are too costly to be used, and surrogate
models have been developed that capture the features of these waveforms at
reduced computational cost~\cite{Purrer:2015tud,Lackey:2016krb}.  

The specific choice of the waveform model $H$ for an analysis depends on the
physics that we wish to explore, computational cost limitations, and the level
of accuracy desired in the model. A variety of waveform models are available
for use in PyCBC Inference, either directly implemented in PyCBC or via calls
to the LIGO Algorithm Library (LAL)~\cite{lal}. We  refer to the PyCBC and LAL
documentation, and references therein, for detailed descriptions of these
models. In this paper, we demonstrate the use of PyCBC Inference using the
IMRPhenomPv2~\cite{Schmidt:2014iyl,Hannam:2013oca} waveform model  for binary
black  hole mergers. This model
captures the inspiral-merger-ringdown physics of spinning, precessing binaries and
parameterizing spin effects using a spin magnitude $a_{j}$, an azimuthal angle
$\theta_{j}^{a}$, and a polar angle $\theta_{j}^{p}$ for each of the two
compact objects. Examples of using PyCBC Inference with different waveform models include the analysis of
Ref.~\cite{De:2018uhw} that used the TaylorF2 post-Newtonian waveform model with
tidal corrections, and Ref.~\cite{PhysRevD.97.124069} that used a ringdown-only
waveform that models the quasi-normal modes of the remnant black hole.

\subsection{Likelihood Function}\label{sec:likelihood}

The data observed by the gravitational-wave detector network enters Bayes'
theorem through the likelihood $\likelihood$ in Eq.~(\ref{eqn:bayes}).
Currently, PyCBC Inference assumes that the each detector produces stationary,
Gaussian noise $n_{i}(t)$ that is uncorrelated between the detectors in the
network. The observed data is then $d_{i}(t) = n_{i}(t) + s_{i}(t)$, where
$s_i(t)$ is the gravitational waveform observed in the $i$-th detector.  For
detectors that are \emph{not} identical and co-located (as in the case of the
LIGO-Virgo network), each detector observers a slightly different waveform due
to their different antennae patterns, however the signal in the $i$-th
detector can be calculated given the subset of the parameters $\vec{\pset}$
that describes the location of the binary.

Under these assumptions, the appropriate form of $\likelihood$ is the
well-known likelihood for a signal of known morphology in Gaussian noise (see
e.g.  Ref.~\cite{Wainstein:1962} for its derivation), which is
given by
\begin{eqnarray}\label{eqn:log_likelihood}
\likelihood = \exp \left[ -\frac{1}{2} \sum_{i = 1}^{N} \left<\tilde{n}_{i}(f) | \tilde{n}_{i}(f)\right> \right] \nonumber\\ 
= \exp \left[ -\frac{1}{2} \sum_{i = 1}^{N} \left<\tilde{d}_{i}(f) - \tilde{s}_{i}(f, \vec{\pset})| \tilde{d}_{i}(f) - \tilde{s}_{i}(f, \vec{\pset})\right> \right] ,
\end{eqnarray}
where $N$ is the number of detectors in the network.
The inner product $\langle\tilde{a} | \tilde{b}\rangle$ is
\begin{equation}
\left<\tilde{a}_i(f) | \tilde{b}_i(f)\right> = 4\Re \int_{0}^{\infty} \frac{\tilde{a}_i(f) \tilde{b}_i(f)}{S^{(i)}_n(f)} \mathrm{d}f \,,
\end{equation}
where $S^{(i)}_n(f)$ is the power spectral density of the of the $i$-th
detector's noise.  Here, $\tilde{d}_{i}(f)$ and $\tilde{n}_{i}(f)$ are the
frequency-domain representations of the data and noise, obtained by a Fourier
transformation of $d_{i}(t)$ and $n_{i}(t)$, respectively.  The model waveform
$\tilde{s}_{i}(f, \vec{\pset})$ may be computed directly in the frequency
domain, or in the time domain and then Fourier transformed to the frequency
domain.  There are several operations (e.g. Fourier transforms, noise power
spectral density estimation, and inner products) that are common between the
calculation of Eq.~(\ref{eqn:log_likelihood}) and the computation of the matched
filter signal-to-noise ratio (SNR) in PyCBC~\cite{Allen:2005fk,Usman:2015kfa,Nitz:2017svb}. PyCBC
Inference uses these existing functions, where appropriate.

In general, gravitational-wave signals consist of a superposition of harmonic
modes. However, in many cases it is sufficient to model only the most dominant
mode, since the sub-dominant harmonics are too weak to be measured. In this case,
the signal observed in all detectors has the same simple dependence on the fiducial phase $\phi$,
\begin{equation}
\tilde{s}_i(f, \vec{\pset}, \phi) = \tilde{s}_i^0(f, \vec{\pset}, 0) e^{i\phi}.
\end{equation}
The posterior probability $\posterior$ can be analytically marginalized over $\phi$
for such models~\cite{Wainstein:1962}. Assuming a uniform prior on
$\phi \in [0,2\pi)$, the marginalized posterior is
\begin{eqnarray}
\label{eqn:marginalized_phase}
\log \posterior &\propto \log \prior +
        I_0\left(\left|\sum_i O(\tilde{s}^0_i, \tilde{d}_i)\right|\right) \nonumber \\
        & \qquad - \frac{1}{2}\sum_i\left[ \left<\tilde{s}^0_i, \tilde{s}^0_i\right> -
                                \left<\tilde{d}_i, \tilde{d}_i\right> \right],
\end{eqnarray}
where
\begin{equation*}
\tilde{s}_i^0 \equiv \tilde{s}_i(f, \vec{\pset}, \phi=0),
\end{equation*}
\begin{equation*}
O(\tilde{s}^0_i, \tilde{d}_i) \equiv 4 \int_0^\infty
        \frac{\tilde{s}_i^*(f; \pset,0)\tilde{d}_i(f)}{S^{(i)}_n(f)}\mathrm{d}f,
\end{equation*}
and $I_0$ is the modified Bessel function of the first kind.

We have found that analytically
marginalizing over $\phi$ in this manner reduces the computational cost of
the analysis by a factor of 2 -- 3. The
IMRPhenomPv2 model that we use here is a simplified model of precession that
allows for this analytic marginalization~\cite{Schmidt:2014iyl,Hannam:2013oca}. 
Since fiducial phase is generally a
nuisance parameter, we use this form of the likelihood function in
Secs.~\ref{sec:simpop} and \ref{sec:astro_events}.

\subsection{Sampling Methods}\label{sec:sampling}

Stochastic sampling techniques, and in particular Markov-chain Monte Carlo (MCMC) methods~\cite{metropolis1953,geman:1984,Gilks:1996,Gelman95bayesiandata}, have been used to numerically sample the posterior probability density function of astrophysical parameters for binary-merger signals~\cite{Christensen:2001cr,Christensen:2004bc,Rover:2006ni,Rover:2006bb}.
Ensemble MCMC algorithms use multiple Markov chains to sample the parameter space.
A simple choice to initialize the $k$-th Markov chain in the ensemble is to draw a set of parameters $\vec{\pset}_{1}^{(k)}$ from the prior probability density function.
The Markov chains move around the parameter space according to the following set of rules.
At iteration $l$, the $k$-th Markov chain has the set of parameters $\vec{\pset}_{l}^{(k)}$.
The sampling algorithm chooses a new proposed set of parameters $\vec{\pset}_{l'}^{(k)}$ with probability $Q(\vec{\pset}_{l}^{(k)}, \vec{\pset}_{l'}^{(k)})$.
When a new set of parameters is proposed, the sampler computes an acceptance probability $\gamma$ which determines if the Markov chain should move to the proposed parameter set $\vec{\pset}_{l'}^{(k)}$ such that $\vec{\pset}_{l+1}^{(k)} = \vec{\pset}_{l'}^{(k)}$.
If $\vec{\pset}_{l'}^{(i)}$ is rejected, then $\vec{\pset}_{l+1}^{(k)} = \vec{\pset}_{l}^{(k)}$.
After a sufficient number of iterations, the ensemble converges to a distribution that is proportional to a sampling of the posterior probability density function.
The true astrophysical parameters $\vec{\pset}$ can then be estimated from histograms of the position of the Markov chains in the parameter space.
Different ensemble sampling algorithms make particular choices for the proposal probability $Q(\vec{\pset}_{l}^{(k)}, \vec{\pset}_{l'}^{(k)})$ and acceptance probability $\gamma$.

The open-source community has several well-developed software packages that implement algorithms for sampling the posterior probability density function.
PyCBC Inference leverages these developments, and we have designed a flexible framework that allows the user to choose from multiple ensemble sampling algorithms. Currently, PyCBC Inference supports the open-source ensemble sampler \texttt{emcee}~\cite{emcee,emcee_repo}, its parallel-tempered version \texttt{emcee\_pt}~\cite{ptemcee_repo,vousden:2016}, and the \texttt{kombine}~\cite{kombine,kombine_repo} sampler.
All of the three are ensemble MCMC samplers.
The sampling algorithm advances the positions of the walkers based on their previous positions and provides PyCBC Inference the positions of the walkers along the Markov chain.

The \texttt{emcee\_pt} sampler is a parallel-tempered sampler which advances multiple ensembles based on the tempering or the ``temperatures'' used to explore the posterior probability density function.
The posterior probability density function for a particular temperature $T$ is modified such that
\begin{equation}
p_{T}(\vec{\pset}|\vec{d}(t),H) = \frac{ \likelihood^{\frac{1}{T}} \prior }{p(\vec{d}(t)|H)} .
\end{equation}
The \texttt{emcee\_pt} sampler uses several temperatures in parallel, and the position of Markov chains are swapped between temperatures using an acceptance criteria described in Ref.~\cite{vousden:2016}.
Mixing of Markov chains from the different temperatures makes parallel-tempered samplers suitable for sampling posterior probability density functions with widely separated modes in the parameter space~\cite{vousden:2016}.
The \texttt{emcee} sampler performs the sampling using one temperature where $T = 1$.

The \texttt{kombine} sampler on the other hand uses clustered kernel-density estimates to construct its proposal distribution, and proposals are accepted using the Metropolis--Hastings condition~\cite{Hastings:1970aa}.
The \texttt{kombine} sampler has been included in PyCBC Inference due to its efficient sampling which significantly lowers the computational cost of an analysis relative to the \texttt{emcee\_pt} sampler. However, in Sec.~\ref{sec:simpop}, we found that the nominal configuration of the \texttt{kombine} sampler produced biased estimates of parameters for binary black holes.

\subsection{Selection of Independent Samples}\label{sec:select}

The output returned by the sampling algorithms discussed in Sec.~\ref{sec:sampling} are Markov chains. Successive states of these chains are not independent, as Markov processes depend on the previous state~\cite{Christensen:2004jm}.
The autocorrelation length $\tau_K$ of a Markov chain is a measure of the number of iterations required to produce independent samples of the posterior probability density function~\cite{Madras:1988ei}.
The autocorrelation length of the $k$-th Markov chain $X_{l}^{(k)} = \{ \vec{\pset}_{g}^{(k)} ; 1 < g < l \}$ of length $l$ obtained from the sampling algorithm is defined as 
\begin{equation}
\tau_K = 1 + 2 \sum_{i=1}^{K} \hat{R}_{i},
\end{equation}
where $K$ is the first iteration along the Markov chain the condition $m \tau_K \leq K$ is true, $m$ being a parameter which in PyCBC  Inference is set to $5$~\cite{Madras:1988ei}.
The autocorrelation function $\hat{R}_{i}$ is defined as
\begin{equation}
\hat{R}_{i} = \frac{1}{l \sigma^{2}} \sum_{t=1}^{l-i} \left( X_{t} - \mu \right) \left( X_{t+i} - \mu \right),
\end{equation}
where $X_t$ are the samples of $X_{l}^{(k)}$ between the 0-th and the $t$-th iteration, $X_{t+i}$ are the samples of $X_{l}^{(k)}$ between the 0-th and the $(t+1)$-th iterations.
Here, $\mu$ and $\sigma^2$ are the mean and variance of $X_t$, respectively.

The initial positions of the Markov chains influence their subsequent positions. 
The length of the Markov chains before they are considered to have lost any memory of the initial positions is called the ``burn-in'' period.
It is a common practice in MCMC analyses to discard samples from the burn-in period to prevent any bias introduced by the initial positions of the Markov chains on the estimates of the parameters from the MCMC.
PyCBC Inference has several methods to determine when the Markov chains are past the burn-in period.
Here, we describe two methods, \texttt{max\_posterior} and \texttt{n\_acl}, which we have found to work well with the \texttt{kombine} and \texttt{emcee\_pt} samplers used in Sections \ref{sec:simpop} and \ref{sec:astro_events}.

The \texttt{max\_posterior} algorithm is an implementation of the burn-in test used for the MCMC sampler in Ref.~\cite{Veitch:2014wba}. In this method, the $k$-th Markov chain is considered to be past the burn-in period at the first iteration $l$ for which
\begin{equation}
\log \mathcal{L}_{l}^{(k)} \geq \max_{k,l} \log \mathcal{L} - \frac{N_p}{2} ,
\end{equation}
where $\mathcal{L}$ is the prior-weighted likelihood
\begin{equation}
\mathcal{L} = \likelihood \prior ,
\end{equation}
and $N_p$ is the number of dimensions in the parameter space. The maximization $\max_{k,l} \log \mathcal{L}$ is carried out over all Markov chains and iterations. The ensemble is considered to be past the burn-in period at the first iteration where all chains pass this test.
We have found this test works well with the \texttt{kombine} sampler if the network signal-to-noise ratio of the signal is $\gtrsim 5$. 

While the \texttt{max\_posterior} test works well with the \texttt{kombine} sampler, we have found that it underestimates the burn-in period when used with the \texttt{emcee\_pt} sampler.
Instead we use the \texttt{n\_acl} test with the \texttt{emcee\_pt} sampler.
This test posits that the sampler is past the burn-in period if the length of the chains exceed $10$ times the autocorrelation length.
The autocorrelation length is calculated using samples from the second half of the Markov chains.
If the test is satisfied, the sampler is considered to be past the burn-in period at the midway point of the Markov chains.

Correlations between the neighboring samples after the burn-in period are removed by ``thinning'' or drawing samples from the Markov chains with an interval of the autocorrelation length~\cite{Christensen:2004jm}. This is done so that the samples used to estimate the posterior probability density function are independent. Therefore, the number of independent samples of the posterior probability density function is equal to the number of Markov chains used in the ensemble times the number of iterations after the burn-in period divided by the autocorrelation length. PyCBC Inference will run until it has obtained the desired number of independent samples after the burn-in period.

\subsection{Credible Intervals}\label{sec:ints}

After discarding samples from the burn-in period and thinning the remaining samples of the Markov chains, the product is the set of independent samples as described in Sec.~\ref{sec:select}.
Typically we summarize the measurement of a given parameter using a credible interval.
The $x$\% credible interval is an interval where the true parameter value lies with a probability of $x$\%.
PyCBC Inference provides the capability to calculate credible intervals based on percentile values.
In the percentile method, the $x$\% credible interval of a parameter value is written as $A_{- B}^{+ C}$ where $A$ is typically the 50-th percentile (median) of the marginalized histograms.
The values $A - B$ and $A + C$ represent the lower and upper boundaries of the $x$-th percentile respectively.

An alternative method of calculating a credible interval estimate is the Highest Posterior Density (HPD) method.
An $x\%$ HPD interval is the shortest interval that contains $x\%$ of the probability.
The percentile method explained above imposes a non-zero lower boundary to the interval being measured.
This can be perceived as a limitation in cases where the weight of histogram at the $\sim$ 0-th percentile is not significantly different from the weight at the lower boundary of the credible interval.
Intervals constructed using the HPD method may be preferred in such cases.
Previous studies have noted that HPD intervals may be useful when the posterior distribution is not symmetric~\cite{Chen2000}.
PyCBC Inference uses HPD to construct confidence contours for two-dimensional marginal distributions, but HPD is not used in the contruction of one-dimensional credible intervals for a single parameter. This functionality will be included in a future release of PyCBC Inference.

\section{The PyCBC Inference Toolkit}\label{sec:code}
\begin{landscape}
\begin{figure}
\includegraphics[width=\linewidth]{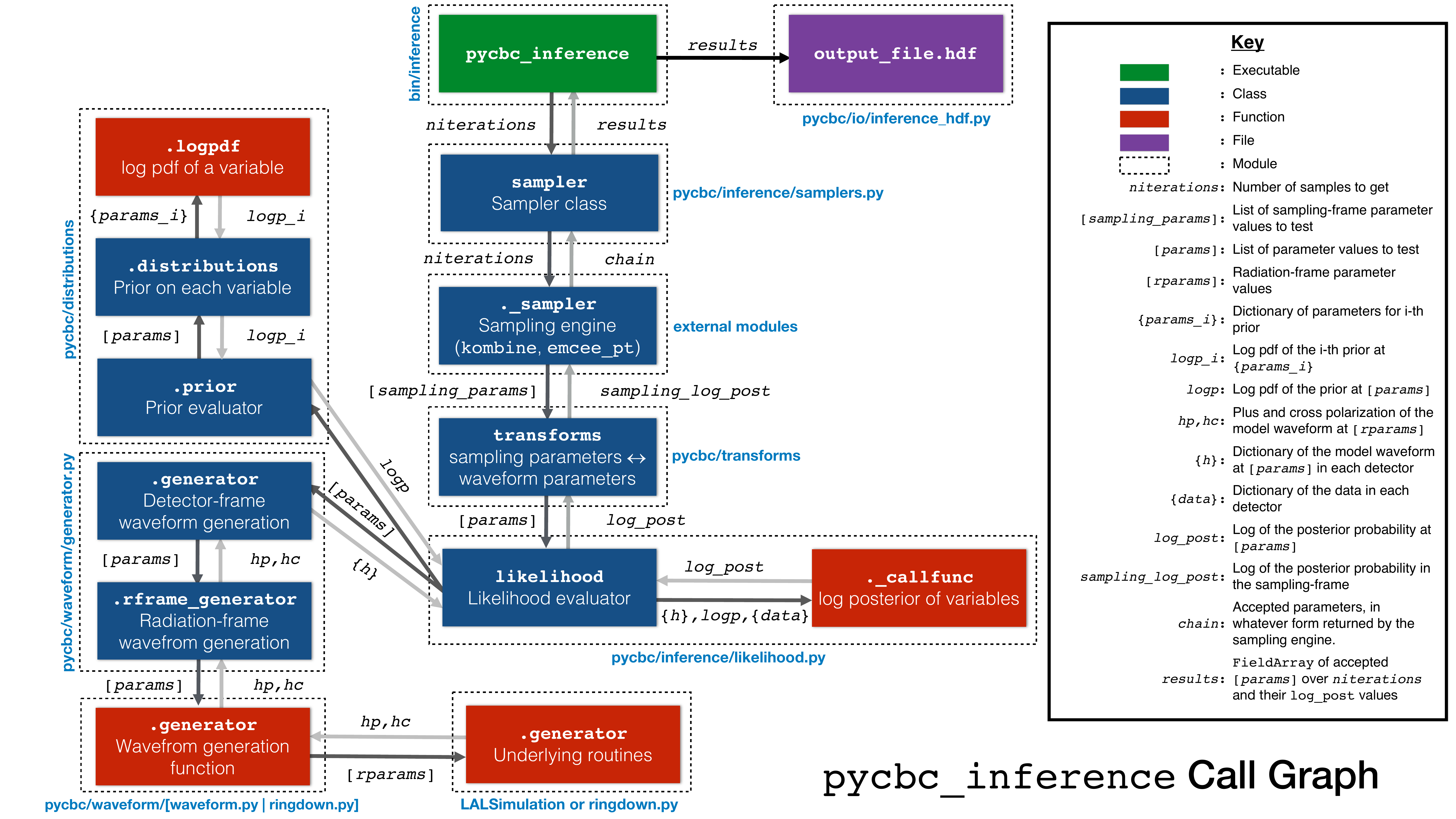}
\centering
\caption{The executable \texttt{pycbc\_inference} samples the posterior probability density function.
For an iteration in an ensemble MCMC algorithm, the \texttt{Sampler} object uses the \texttt{LikelihoodEvaluator} object to compute the natural logarithm of the posterior probability, and returns it to \texttt{pycbc\_inference}. The \texttt{LikelihoodEvaluator} object uses the \texttt{Generator} object to generate the waveform and \texttt{Distribution} objects to evaluate the prior probability density function.
Samples are periodically written to the output file.}
\label{fig:callgraph}
\end{figure}
\end{landscape}

In this section we describe the implementation of PyCBC Inference within the
broader PyCBC toolkit. PyCBC provides both modules for developing code and
executables for performing specific tasks with these modules. The code is
available on the public GitHub repository at
\url{https://github.com/gwastro/pycbc}, with executables located in the
directory \texttt{bin/inference} and the modules in the directory
\texttt{pycbc/inference}.  PyCBC Inference provides an executable called
\texttt{pycbc\_inference} that is the main engine for performing Bayesian
inference with PyCBC.
A call graph of \texttt{pycbc\_inference} is shown in
Figure~\ref{fig:callgraph}. In this section, we review the structure
of the main engine and the
Python objects used to build \texttt{pycbc\_inference}.

\subsection{\texttt{pycbc\_inference} executable}

The methods presented in Sec.~\ref{sec:models}, \ref{sec:likelihood}, \ref{sec:sampling}, \ref{sec:select}, and~\ref{sec:ints} are used to build the executable \texttt{pycbc\_inference}.
For faster performances, \texttt{pycbc\_inference} can be run on high-throughput computing frameworks such as HTCondor~\cite{beowulfbook-condor,condor-practice} and the processes for running the sampler can be parallelized over multiple compute nodes using MPI~\cite{dalcin2011parallel,Dalcin:2008,Dalcin:2005:MP:1196218.1196228}.
The execution of the likelihood computation and the PSD estimation are done using either single-threaded or parallel FFT engines, such as FFTW~\cite{FFTW05} or the Intel Math Kernel Library (MKL).
For maximum flexibility in heterogeneous computing environments, the processing scheme to be used is specified at runtime as a command line option to \texttt{pycbc\_inference}.

The input to \texttt{pycbc\_inference} is a configuration file which contains up to seven types of sections.
The \texttt{variable\_args} section specifies the parameters that are to be varied in the MCMC.
There is a \texttt{prior} section for each parameter in the \texttt{variable\_args} section 
which contains arguments to initialize the prior probability density function for that parameter.
There is a \texttt{static\_args} section specifying any parameter for waveform generation along with its assigned value that should be fixed in the ensemble MCMC.
Optionally, the configuration file may also include a \texttt{constraint} section(s) containing any conditions that constrain the prior probability density functions of the parameters.
For efficient convergence of a Markov chain, it may be desirable to sample the prior probability density function in a different coordinate system than the parameters defined in the \texttt{variable\_args} section or the parameters inputted to the waveform generation functions.
Therefore, the configuration file may contain a \texttt{sampling\_parameters} and \texttt{sampling\_transform} section(s) that specifies the transformations between parameters in the \texttt{variable\_args} sections and the parameters evaluated in the prior probability density function.
Finally, the waveform generation functions recognize only a specific set of input parameters.
The \texttt{waveform\_transforms} section(s) may be provided which maps parameters in the \texttt{variable\_args} section to parameters understood by the waveform generation functions.
More details on application of constraints and execution of coordinate transformations are provided in Sec.~\ref{subsec:transform} and \ref{subsec:likelihoodeval} respectively.

{
The location of the configuration file, gravitational-wave detector data files, data conditioning settings, and settings for the ensemble MCMC are supplied on the command line interface to \texttt{pycbc\_inference}.
The results from running \texttt{pycbc\_inference} are stored in a
\unskip\parfillskip 0pt \par}
\noindent
HDF~\cite{andrew_collette_2018_1246321} file whose location is provided on the command line to \texttt{pycbc\_inference} as well.
The main results of interest are stored under the HDF groups \texttt{[`samples']} and \texttt{[`likelihood\_stats']}. The \texttt{[`samples']} group contains the history of the Markov chains as separate datasets for each of the variable parameters.
The \texttt{[`likelihood\_stats']} group contains a dataset of the natural logarithm of the Jacobian which is needed to transform from the variable parameters to sampling parameters, a dataset containing natural logarithm of the likelihood ratio $\log \likelihood/ p(\vec{d}(t)|\vec{n})$ and a dataset containing the natural logarithm of the prior probabilities.
The natural logarithm of the noise likelihood $\log p(\vec{d}(t)|\vec{n})$ is stored as an attribute in the output file, and the likelihood is the summation of this quantity with the natural logarithm of the likelihood ratio.
Each of the datasets under the \texttt{[`samples']} group and the \texttt{[`likelihood\_stats']} group has shape \texttt{nwalkers} $\times$ \texttt{niterations} if the sampling algorithm used in the analysis did not include parallel tempering, and has shape \texttt{ntemps} $\times$ \texttt{nwalkers} $\times$ \texttt{niterations} for parallel-tempered samplers.
Here, \texttt{nwalkers} is the number of Markov chains, \texttt{niterations} is the number of iterations, and \texttt{ntemps} is the number of temperatures.

\texttt{pycbc\_inference} has checkpointing implemented which allows users to resume an analysis from the last set of Markov chains positions written to the output file.
It is computationally expensive to obtain the desired number of independent samples using ensemble MCMC methods, and the \texttt{pycbc\_inference} processes may terminate early due to problems on distributed-computing networks.
Therefore, the samples from the Markov chains should be written at regular intervals so \texttt{pycbc\_inference} can resume the ensemble MCMC from the position of the Markov chains 
near the state the process was terminated.
The frequency \texttt{pycbc\_inference} writes the samples from Markov chains and the state of the random number generator to the output file and a backup file is specified by the user on the command line.
A backup file is written by \texttt{pycbc\_inference} because the output file from \texttt{pycbc\_inference} may be corrupted.
For example, if the  process is aborted while writing to the output file, then the output file may be corrupted.
In that case, samples and the state of the random number generator are loaded from the backup file, and the backup file is copied to the output file.
This ensures that the \texttt{pycbc\_inference} process can always be resumed.

For analyses that use the \texttt{emcee\_pt} sampler, the likelihood can be used to compute the natural logarithm of the evidence using the \texttt{emcee\_pt} sampler's \texttt{thermodynamic\_integration\_log\_evidence} function \cite{ptemcee_repo}.
Then, the evidences from two analyses can be used to compute the Bayes factor $\mathcal{B}$ for the comparison of two waveform models.

We provide example configuration files and run scripts for the analysis of the binary black hole mergers detected in the Advanced LIGO's first observing run in Ref.~\cite{ResultsRepo}.
These examples can be used with the open-source datasets provided by the LIGO Open Science Center~\cite{Vallisneri:2014vxa}.
The results of these analyses are presented in Sec.~\ref{sec:astro_events}.

\subsection{\texttt{Sampler} objects}

The PyCBC Inference modules provide a set of \texttt{Sampler} objects which execute the Bayesian sampling methods.
These objects provide classes and functions for using open-source samplers such as \texttt{emcee}~\cite{emcee_repo}, \texttt{emcee\_pt}~\cite{ptemcee_repo} or \texttt{kombine}~\cite{kombine_repo}.
This acts as an interface between PyCBC Inference and the external sampler package.
The executable \texttt{pycbc\_inference} initializes, executes, and saves the output from the \texttt{Sampler} objects.
A particular \texttt{Sampler} object is chosen on the command line of \texttt{pycbc\_inference} with the \texttt{--sampler} option. The \texttt{Sampler} object provides the external sampler package the positions of the walkers in the parameter space, the natural logarithm of the posterior probabilities at the current iteration, the current ``state'' determined from a random number generator, and the number of iterations that the sampler is requested to run starting from the current iteration.
After running for the given number of iterations, the sampler returns the updated positions of the Markov chains, the natural logarithm of the posterior probabilities, and the new state. 

\subsection{\texttt{Transform} objects} \label{subsec:transform}

The \texttt{Transform} objects in PyCBC Inference are used to perform transformations between different coordinate systems.
Currently, the \texttt{Transform} objects are used in two cases: sampling transforms and waveform transforms. 

Sampling transforms are used for transforming parameters that are varied in the
ensemble MCMC to a different coordinate system before evaluating the prior
probaility density function.  Since there exists degeneracies between several
parameters in a waveform model it is useful to parameterize the waveform using
a preferred set of parameters which could minimize the correlations.  This
leads to more efficient sampling, and therefore, it leads to faster convergence
of the Markov chains.  One example of a sampling transformation is the
transformation between the component masses $m_1$ and $m_2$ to chirp mass
$\mathcal{M}$ and mass ratio $q$. The convention adopted for $q$ in PyCBC
Inference is $q = m_1 / m_2$, where $m_1$ and $m_2$ are the component masses
with $m_1 > m_2$. The chirp mass $\mathcal{M}$
is the most accurately measured parameter in a waveform model because it is in
the leading order term of the post-Newtonian expression of the waveform model.
In contrast, the degeneracies of the mass ratio with spin introduces
uncertainties in measurements of the component masses.  Therefore, sampling in
$(\mathcal{M}$ and $q)$ proves to be more efficient than $m_1$ and
$m_2$~\cite{Rover:2006ni,Veitch:2014wba,Farr:2015lna}.  In the GW150914,
LVT151012, and GW151226 configuration files in \cite{ResultsRepo}, we
demonstrate how to allow the \texttt{Sampler} object to provide priors in the
$(m_1, m_2)$ coordinates, and specify sampling transformations to the
$(\mathcal{M}, q)$ coordinates.

Waveform transforms are used to transform any variable parameters in the ensemble MCMC that may not be understood by the waveform model functions.
In PyCBC, the waveform model functions accept the following parameters: component masses $m_1$ and $m_2$, $d_\mathrm{L}$, $\iota$, $t_c$, $\phi$, and any additional spin parameters in Cartesian coordinates. The convention adopted in PyCBC for $\iota$ denotes $\iota = 0$ as a
``face-on'' binary (line of sight parallel to binary angular momentum), $\iota
= \frac{\pi}{2}$ as an ``edge-on'' binary (line of sight perpendicular to
binary angular momentum), and $\iota = \pi$ as a ``face-off'' binary (line of
sight anti-parallel to binary angular momentum).

\subsection{\texttt{LikelihoodEvaluator} object}\label{subsec:likelihoodeval}

The \texttt{LikelihoodEvaluator} object computes the natural logarithm of the prior-weighted likelihood given by the numerator of Eq.~\ref{eqn:bayes}.
Since the evidence is constant for a given waveform model, then the prior-weighted likelihood is proportional to the posterior probability density function, and it can be used in sampling algorithms to compute the acceptance probability $\gamma$ instead of the full posterior probability density function.
The prior-weighted likelihood is computed for each new set of parameters as the \texttt{Sampler} objects advance the Markov chains through the parameter space.

\subsection{\texttt{Distribution} objects}

The \texttt{LikelihoodEvaluator} object must compute the prior probability density function $p(\vec{\pset}|H)$. 
There exists several \texttt{Distribution} objects that provide functions for evaluating the prior probability density function to use for each parameter, and for drawing random samples from these distributions. Currently, PyCBC Inference provides the following \texttt{Distribution}s:
\begin{enumerate}
\item \texttt{Arbitrary} : Reads a set of samples stored in a HDF format file and uses Gaussian kernel-density estimation~\cite{scipy_kde} to construct the distribution.
\item \texttt{CosAngle} : A cosine distribution.
\item \texttt{SinAngle} : A sine distribution.
\item \texttt{Gaussian} : A multivariate Gaussian distribution.
\item \texttt{Uniform} : A multidimensional uniform distribution.
\item \texttt{UniformAngle} : A uniform distribution between 0 and 2$\pi$.
\item \texttt{UniformLog} : A multidimensional distribution that is uniform in its logarithm.
\item \texttt{UniformPowerLaw} : A multidimensional distribution that is uniform in a power law.
\item \texttt{UniformSky} : A two-dimensional isotropic distribution.
\item \texttt{UniformSolidAngle} : A two-dimensional distribution that is uniform in solid angle.
\end{enumerate}
Multiple \texttt{Distribution} objects are needed to define the prior probability density function for all parameters.
The \texttt{JointDistribution} object combines the individual prior probability density functions, providing a single interface for the \texttt{LikelihoodEvaluator} to evaluate the prior probability density function for all parameters.
As the sampling algorithm advances the positions of the Markov chains, the \texttt{JointDistribution} computes the product of the prior probability density functions for the proposed new set of points in the parameter space.
The \texttt{JointDistribution} can apply additional constraints on the prior probability density functions of parameters and it renormalizes the prior probability density function accordingly.
If multiple constraints are provided, then the union of all constraints are applied.
We demonstrate how to apply a cut on $\mathcal{M}$ and $q$ obtained from the $m_1$ and $m_2$ prior probability density functions in the GW151226 configuration file in Ref.~\cite{ResultsRepo}.

\subsection{\texttt{Generator} objects}

As part of the likelihood calculation described in Sec.~\ref{sec:likelihood}, a waveform $\tilde{s}_{i}(f, \vec{\pset})$ is generated from a waveform model $H$ and set of parameters $\vec{\pset}$.
PyCBC Inference provides \texttt{Generator} objects that allow waveforms $\tilde{s}_{i}(f, \vec{\pset})$ to be generated for waveform models described in Sec.~\ref{sec:models} using PyCBC's interface to LAL~\cite{lal}.
There are also \texttt{Generator} objects provided for generating ringdown waveforms as used in Ref.~\cite{PhysRevD.97.124069}.
Given the waveform model provided in the configuration file, \texttt{pycbc\_inference} will automatically select the associated \texttt{Generator} object.

\section{Validation of the Toolkit}
\label{sec:validation}

\begin{figure*}
\subfloat{\includegraphics[width=0.5\textwidth]{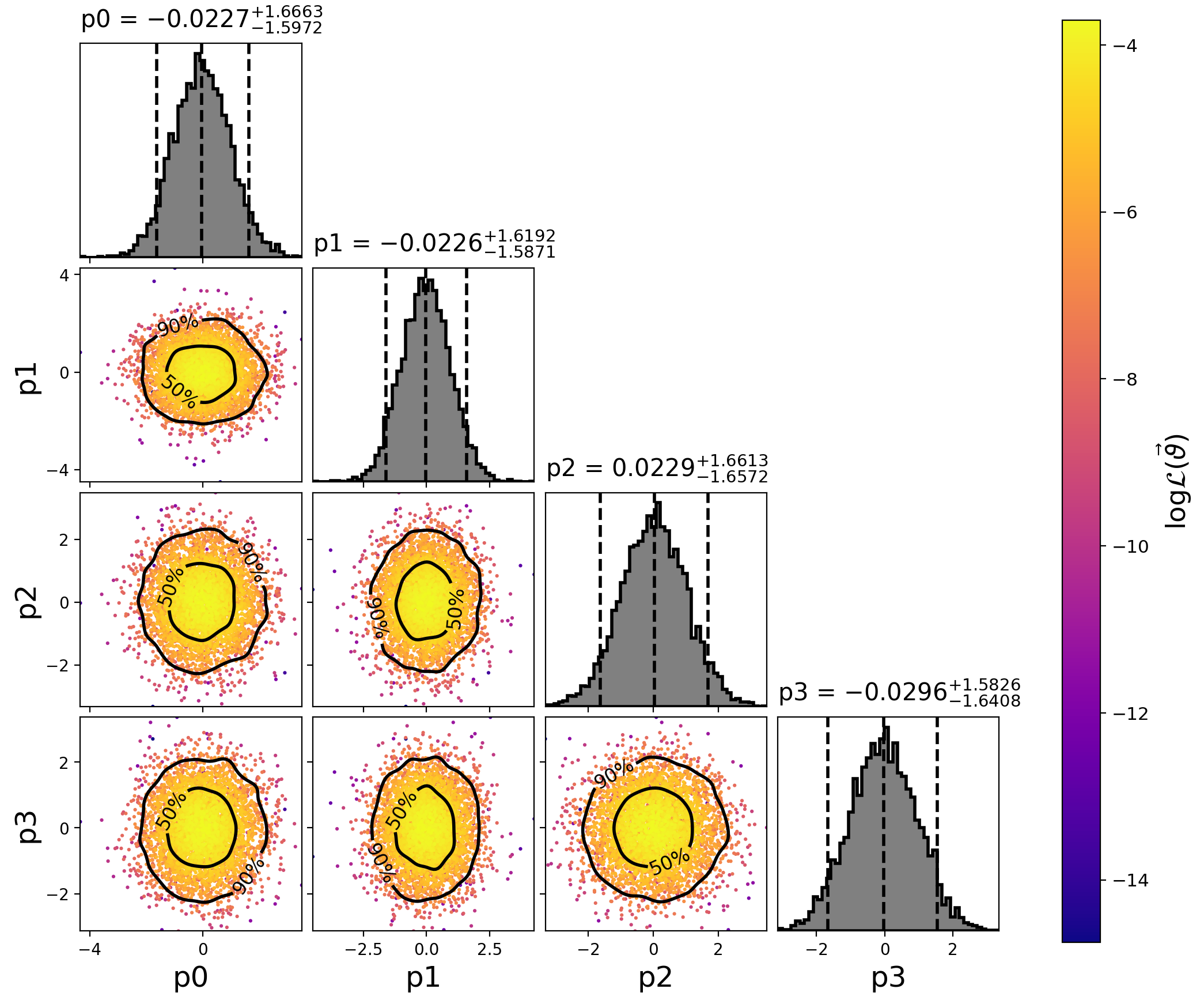}}
\subfloat{\includegraphics[width=0.5\textwidth]{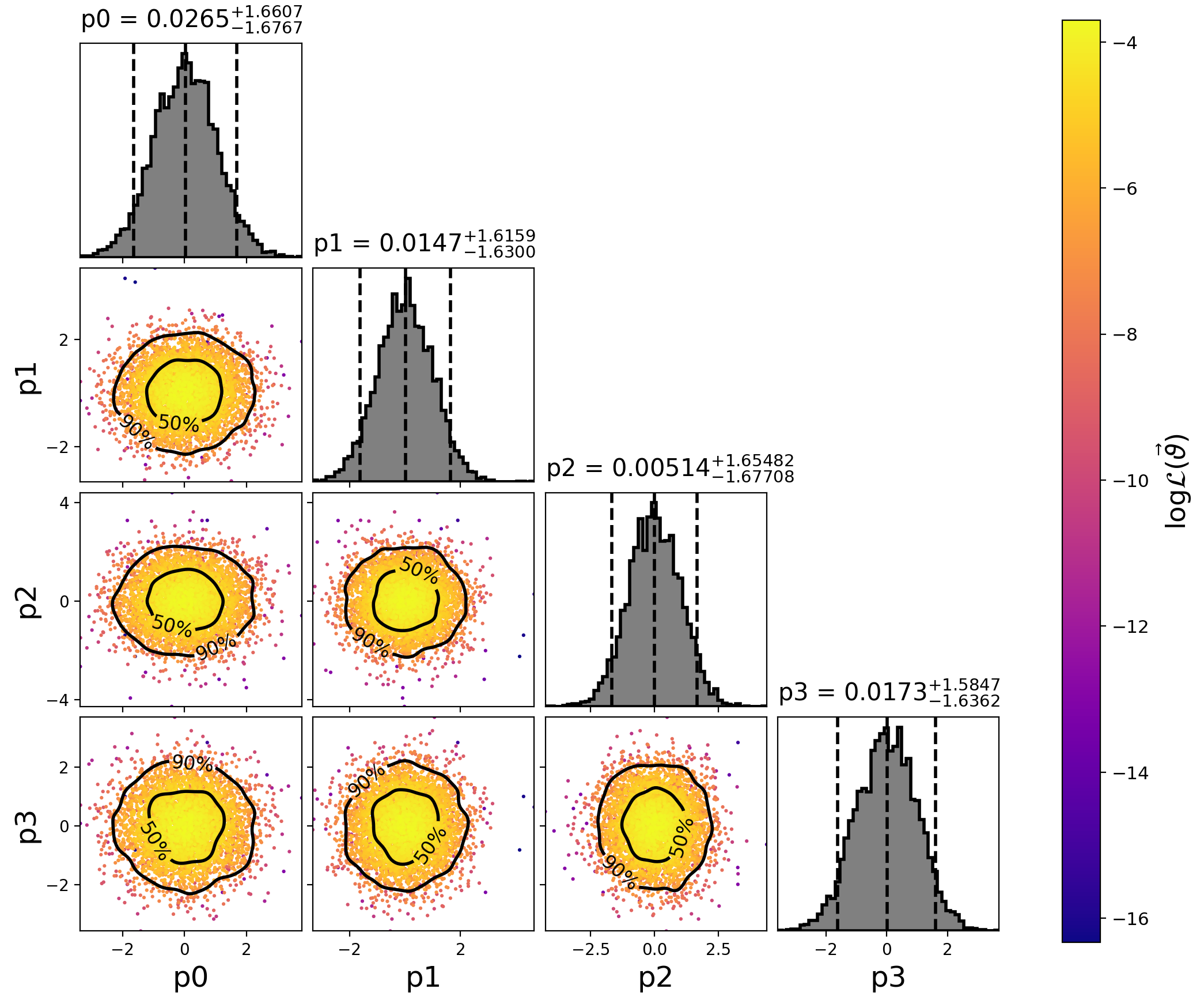}}\caption{\label{fig:Post4dNormal}
The samples of the posterior probability density function for a
four-dimensional normal distribution.  Typically, these results are shown as a
scatter-plot matrix of independent samples.  Here, the points in the
scatter-plot matrix are colored by the natural logarithm of the prior-weighted
likelihood $\log \mathcal{L}(\vec{\pset})$.  At the top of each column is the
marginalized one-dimensional histogram for a particular model parameter.  In
this case, each parameter $p_{i}$ is the mean of a Gaussian in the range (0,
1).  The median and 90\% credible interval are superimposed on the
marginalized histograms.  \emph{Left}: Results obtained from the
\texttt{emcee\_pt} sampler. \emph{Right}: Results obtained from the
\texttt{kombine} sampler.}
\end{figure*}

PyCBC Inference includes tools for visualizing the results of parameter
estimation, and several analytic functions that can be used to test the
generation of known posterior probabilities.  Two common ways to visualize
results are a scatter plot matrix of the independent samples of the Markov
chains, and a marginalized one-dimensional histograms showing the bounds of
each parameter's credible interval.  Analytic likelihood functions available
to validate the code include: the multivariate normal, Rosenbrock, eggbox, and
volcano functions. An example showing the visualization of results from the
multi-variate Gaussian test is shown in Fig.~\ref{fig:Post4dNormal}.  This
figure was generated using the executable
\texttt{pycbc\_inference\_plot\_posterior} which make extensive use of tools
from the open-source packages Matplotlib~\cite{Hunter:2007} and
SciPy~\cite{scipy}. 

We can also  validate the performance of PyCBC Inference by: (i) determining
if the inferred parameters of a population of simulated signals agrees with
known the parameters that  population, and (ii) comparing PyCBC Inference's
parameter credible intervals astrophysical signals  to the published
LIGO-Virgo results that used a different inference code. In this section, we
first check that the credible intervals match the probability of finding the
simulated signal parameters in that interval, that is, that $x\%$ of signals
should have parameter values in the $x\%$ credible interval. We then compare
the recovered parameters of the binary black hole mergers GW150914, GW151226,
and LVT151012 to those published in Ref.~\cite{TheLIGOScientific:2016pea}.

\subsection{Simulated Signals}
\label{sec:simpop}

To test the performance of PyCBC Inference, we generate 100 realizations of stationary Gaussian noise colored by power-spectral densities representative of the sensitivity of Advanced LIGO detectors at the time of the detection of GW150914~\cite{Vallisneri:2014vxa}.
To each realization of noise we add a simulated signal whose parameters were drawn from the same prior probability density function used in the analysis of GW150914~\cite{TheLIGOScientific:2016wfe}, with an additional cut placed on distance to avoid having too many injections with low matched-filter SNR.
The resulting injections have matched-filter SNRs between $5$ and $160$, with the majority between $\sim10$ and $\sim40$.
We then perform a parameter estimation analysis on each signal to obtain credible intervals on all parameters.

We perform this test using both the \texttt{emcee\_pt} and \texttt{kombine} samplers.
For the \texttt{emcee\_pt} sampler we use 200 walkers and 20 temperatures.
We run the sampler until we obtain at least 2000 independent samples after the burn-in period as determined using the \texttt{n\_acl} burn-in test. For the \texttt{kombine} sampler, we use 5000 walkers and the \texttt{max\_posterior} burn-in test.
As a result, we need only to run the \texttt{kombine} sampler until the burn-in test is satisfied, at which point we immediately have 5000 independent samples of the posterior probability density function.

Both simulated signals and the waveforms in the likelihood computation are generated using IMRPhenomPv2~\cite{Schmidt:2014iyl,Hannam:2013oca}.
This waveform model has 15 parameters. To reduce computational cost, we analytically marginalize over the fiducial phase $\phi$ by using Eq.~(\ref{eqn:marginalized_phase}) for the
posterior probability, thereby reducing the number of sampled parameters to 14. For each parameter, we count the number of times the simulated parameter falls within the measured credible interval.

\begin{figure}[t]
\includegraphics[width=\columnwidth]{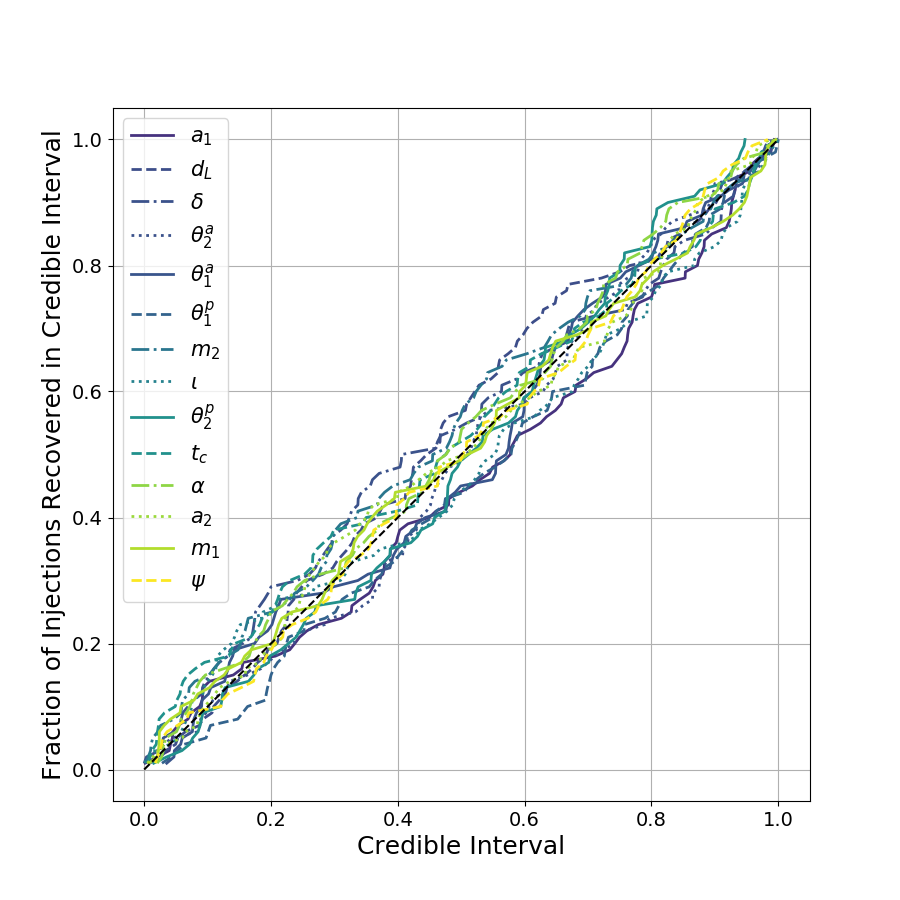}
\includegraphics[width=\columnwidth]{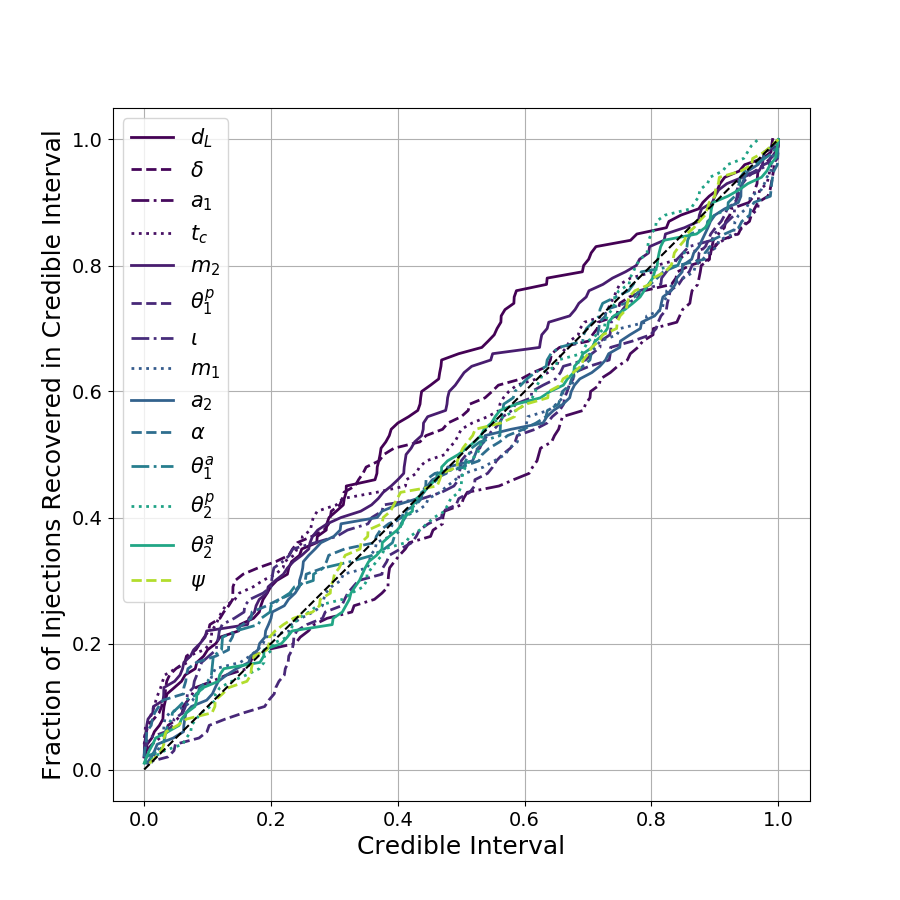}
\caption{Fraction of simulated signals with parameter values within a credible interval as a function of credible interval. Plotted are all 14 parameters varied in the MCMC analyses. The diagonal line indicates the ideal 1-to-1 relation that is expected if the samplers provide unbiased estimates of the parameters. We perform a Kolmogorov-Smirnov (KS) test on each parameter to obtain two-tailed p-value indicating the consistency between the curves and the diagonal line. \textit{Top:} Results using the \texttt{emcee\_pt} sampler. \textit{Bottom:} Results using the \texttt{kombine} sampler.
}
\label{fig:valid}
\end{figure}

Figure~\ref{fig:valid} summarizes the result of this test using the \texttt{emcee\_pt} and \texttt{kombine} samplers. For each of the parameters we plot the fraction of signals whose true parameter value fall within a credible interval as a function of credible interval (this is referred to as a percentile-percentile plot). We expect the former to equal the latter for all parameters, though some fluctuation is expected due to noise. We see that all parameters follow a 1-to-1 relation, though the results from the \texttt{kombine} sampler have greater variance then the \texttt{emcee\_pt} sampler.

To quantify the deviations seen in Fig.~\ref{fig:valid}, we perform a Kolmogorov--Smirnov (KS) test on each parameter to see whether the percentile-percentile curves match the expected 1-to-1 relation.
If the samplers and code are performing as expected, then these p-values should in turn follow a uniform distribution. We therefore perform another KS test on the collection of p-values, obtaining a two-tailed p-value of $0.50$ for \texttt{emcee\_pt} and $0.03$ for \texttt{kombine}. In other words, if \texttt{emcee\_pt} provides an unbiased estimate of the parameters, then there is a $50\%$ chance that we would obtain a collection of percentile-percentile curves more extreme than seen in Fig.~\ref{fig:valid}. For the \texttt{kombine} sampler, the probability of obtaining a more extreme collection of curves than that seen in Fig.~\ref{fig:valid} is only $3\%$.

Based on these results, we conclude that PyCBC Inference does indeed provide unbiased estimates of binary black hole parameters when used with \texttt{emcee\_pt} with the above settings.
The \texttt{kombine} sampler does not appear to provide unbiased parameter estimates when used to sample the full parameter space of precessing binary black holes with the settings we have used.

\subsection{Astrophysical Events}\label{sec:astro_events}
In this section, we present PyCBC Inference measurements of properties of the binary black hole sources of the two gravitational-wave signals GW150914 and GW151226, and the third gravitational-wave signal LVT151012 consistent with the properties of a binary black hole source from Advanced LIGO's first observing run~\cite{Abbott:2016blz,TheLIGOScientific:2016pea}. We perform the parameter estimation analysis on the Advanced LIGO data available for these events at the LIGO Open Science Center~\cite{Vallisneri:2014vxa}. We use the \texttt{emcee\_pt} sampler for these analyses.
For computing the likelihood, we analyze the  gravitational-wave dataset $\vec d(t)$ from the Hanford and Livingston detectors.
$\vec{d}(t)$ in our analyses are taken from GPS time intervals 1126259452 to 1126259468 for GW150914, 1135136340 to 1135136356 for GW151226, and 1128678874 to 1128678906 for LVT151012. Detection of gravitational waves from the search pipeline~\cite{alex_nitz_2018_1208115,Usman:2015kfa,Canton:2014ena,Nitz:2017svb,TheLIGOScientific:2016qqj} gives initial estimates of the mass, and hence estimates of the length of the signal.
From results of the search, LVT151012 was a longer signal with more cycles than the other two events, and LVT151012 had characteristics which were in agreement with a lower mass source than GW150914 and GW151226.
Therefore, more data is required for the analysis of LVT151012.
The PSD used in the likelihood is constructed using the median PSD estimation method described in Ref.~\cite{Allen:2005fk} with 8~s Hann-windowed segments ( overlapped by 4~s ) taken from GPS times 1126258940 to 1126259980 for GW150914, 1135136238 to 1135137278 for GW151226, and 1128678362 to 1128679418 for LVT151012. The PSD estimate is truncated to 4~s in the time-domain using the method described in Ref. \cite{Allen:2005fk}.
The dataset is sampled at 2048~Hz, and the likelihood is evaluated between a low frequency cutoff of 20~Hz and 1024~Hz.

The waveforms $\tilde{s}_{i}(f, \vec{\pset})$ used in the likelihood are generated using the IMRPhenomPv2~\cite{Schmidt:2014iyl,Hannam:2013oca} model implemented in the LIGO Algorithm Library (LAL)~\cite{lal}.
The parameters inferred for these three events are $\vec{\pset}=\{\alpha, \delta, \psi, m_1, m_2, d_L, \iota, t_c, a_1, a_2, \theta_1^a, \theta_2^a, \theta_1^p, \theta_2^p\}$, and we analytically marginalize over the fiducial phase $\phi$.
These parameters form the complete set of parameters to construct a waveform from a binary black hole merger, and are the same parameters that were inferred from the parameter estimation analyses in Ref.~\cite{TheLIGOScientific:2016pea}.  
Since faster convergence of $m_1$ and $m_2$ can be obtained with mass parameterizations of the waveform in $\mathcal{M}$ and $q$ we perform the coordinate transformation from $(m_1, m_2)$ to $(\mathcal{M}, q)$ before evaluating the priors.

We assume uniform prior distributions for the binary component masses $m_{1,2} \in$ [10, 80] M$_\odot$ for GW150914, $m_{1,2} \in$ [5, 80] M$_\odot$ for LVT151012, and $m_{1,2}$ corresponding to chirp mass $\mathcal{M} \in$ [9.5, 10.5] M$_\odot$ and mass ratio $q \in$ [1, 18] for GW151226.
We use uniform priors on the spin magnitudes $a_{1,2} \in$ [0.0, 0.99].
We use a uniform solid angle prior, where $\theta_{1,2}^{a}$ is a uniform distribution $\theta_{1,2}^{a} \in [0, 2\pi)$ and $\theta_{1,2}^{p}$ is a sine-angle distribution.
For the luminosity distance, we use a uniform in volume prior with $d_L \in$ [10, 1000] Mpc for GW150914, $d_L \in$ [10, 1500] Mpc for GW151226, and $d_L \in$ [10, 2500] Mpc for LVT151012.
We use uniform priors for the arrival time $t_c \in [t_s - 0.2~s, t_s + 0.2~s]$ where $t_s$ is the trigger time for the particular event obtained from the gravitational-wave search~\cite{TheLIGOScientific:2016pea,TheLIGOScientific:2016qqj}.
For the sky location parameters, we use a uniform distribution prior for $\alpha \in [0, 2\pi)$ and a cosine-angle distribution prior for $\delta$.
The priors described above are the same as those used in Ref.~\cite{TheLIGOScientific:2016pea} . 

The parameter estimation analysis produces distributions that are a sampling of the posterior probability density function for the variable parameters from the ensemble MCMC.
We map these distributions obtained directly from the analysis to obtain estimates of other parameters of interest such as the chirp mass $\mathcal{M}$, mass ratio $q$, effective spin $\chi_{\mathrm{eff}}$, and the precession spin $\chi_{p}$~\cite{Schmidt:2014iyl} parameters.
We use $d_L$ to relate the detector-frame masses obtained from the MCMC to the source-frame masses using the standard $\Lambda$-CDM cosmology~\cite{Schutz:1986gp,Finn:1992xs}. 

\begin{figure*}[t]
  \includegraphics[width=\textwidth]{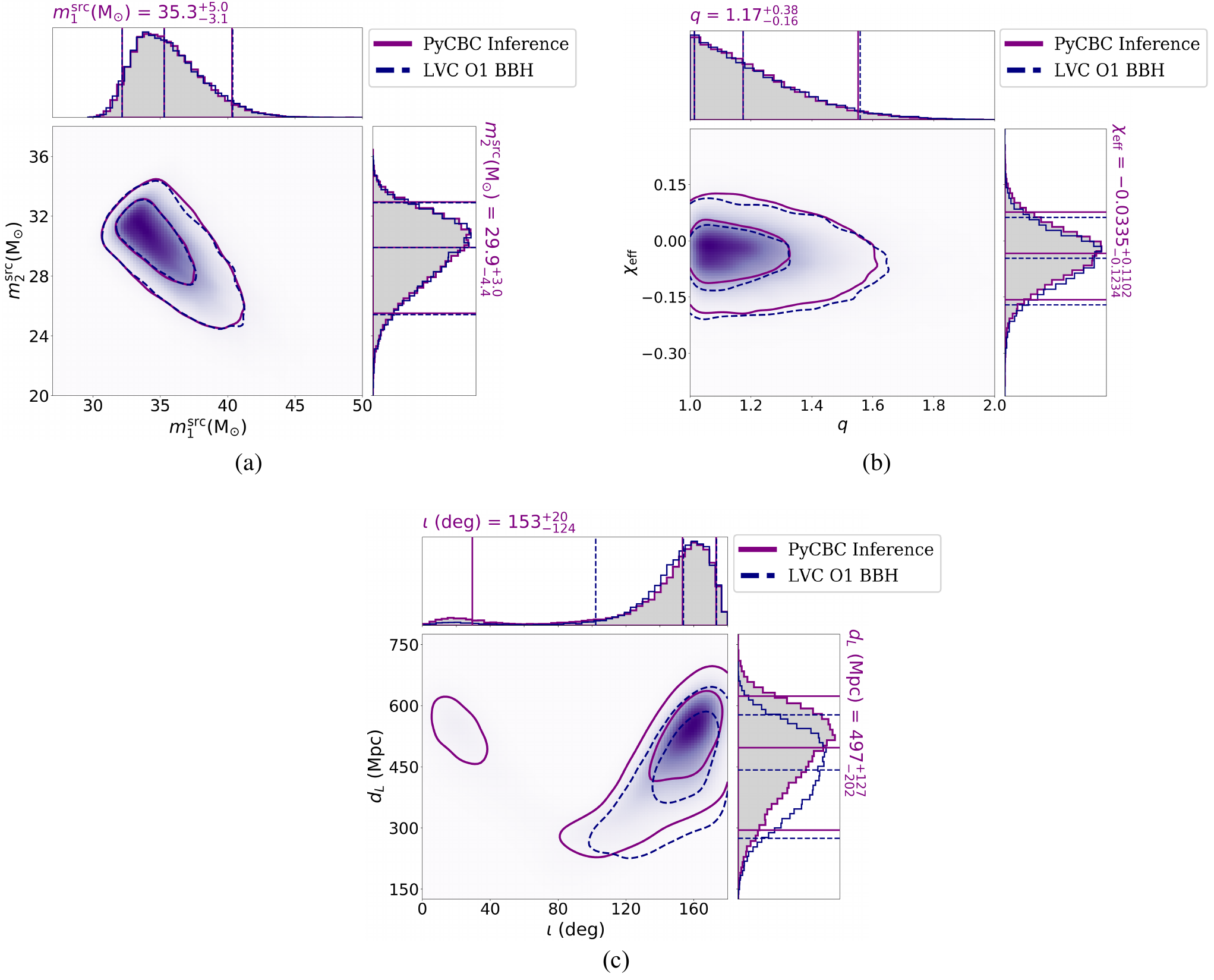}
  \caption{\label{fig:gw150914_plots}Posterior probability densities for the main parameters of interest from the PyCBC Inference analysis of GW150914. The parameters plotted are (a): $m_1^{\mathrm{src}} - m_2^{\mathrm{src}}$, (b): $q - \chi_{\mathrm{eff}}$ (c): $d_{L} - \iota$.  The bottom-left panel in each of (a), (b) and (c) show two-dimensional probability densities with 50\% and 90\% credible contour regions from the PyCBC Inference posteriors. The top-left and the bottom-right panels in each figure show one-dimensional posterior distributions for the individual parameters with solid lines at the 5\%, 50\% and 95\% percentiles. For comparison, we also show 50\% and 90\% credible regions, one-dimensional posterior probabilities with dashed lines at 5\%, 50\% and 95\% percentiles using the posterior samples obtained from the LIGO Open Science Center~\cite{Vallisneri:2014vxa} for the GW150914 analysis reported in Ref.~\cite{TheLIGOScientific:2016pea} using the IMRPhenomPv2 model. The measurements show that masses for GW150914 are much better constrained as compared to the other parameters presented. Though there is support for the system being both ``face-on'' and ``face-off'', there seems to be slightly more preference for a "face-off" system. The posteriors suggest a preference for lower spins. Our measurements are in agreement with the results presented in \cite{TheLIGOScientific:2016pea} within the statistical errors of measurement of the parameters.}
\end{figure*}

\begin{figure*}[t]
  \includegraphics[width=\textwidth]{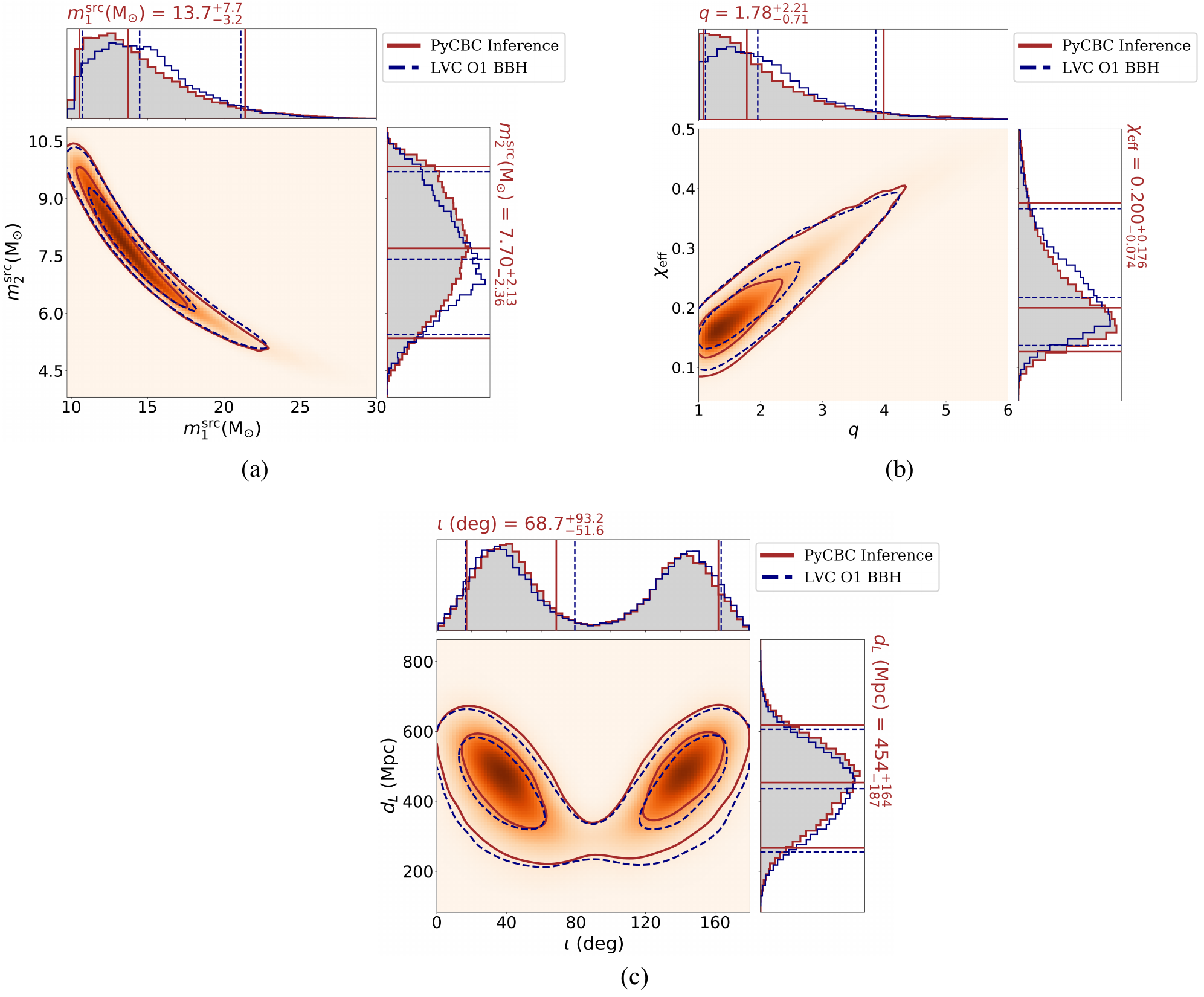}
\caption{\label{fig:gw151226_plots}Posterior probability densities for the main parameters of interest from the PyCBC Inference analysis of GW151226. The parameters plotted are (a): $m_1^{\mathrm{src}} - m_2^{\mathrm{src}}$, (b): $q - \chi_{\mathrm{eff}}$ (c): $d_{L} - \iota$.  The bottom-left panel in each of (a), (b) and (c) show two-dimensional probability densities with 50\% and 90\% credible contour regions from the PyCBC Inference posteriors. The top-left and the bottom-right panels in each figure show one-dimensional posterior distributions for the individual parameters with solid lines at the 5\%, 50\% and 95\% percentiles. For comparison, we also show 50\% and 90\% credible regions, one-dimensional posterior probabilities with dashed lines at 5\%, 50\% and 95\% percentiles using the posterior samples obtained from the LIGO Open Science Center~\cite{Vallisneri:2014vxa} for the GW151226 analysis reported in Ref.~\cite{TheLIGOScientific:2016pea} using the IMRPhenomPv2 model. The measurements show that GW151226 is the lowest mass and fastest spinning binary among the three O1 events presented in this work. The posteriors support asymmetric mass ratios. Inclination $\iota$ and distance $d_L$ are not well constrained, and there is support for the system being both ``face-on'' and ``face-off''. Our measurements are in agreement with the results presented in \cite{TheLIGOScientific:2016pea} within the statistical errors of measurement of the parameters.}
\end{figure*}

\begin{figure*}[t]
  \includegraphics[width=\textwidth]{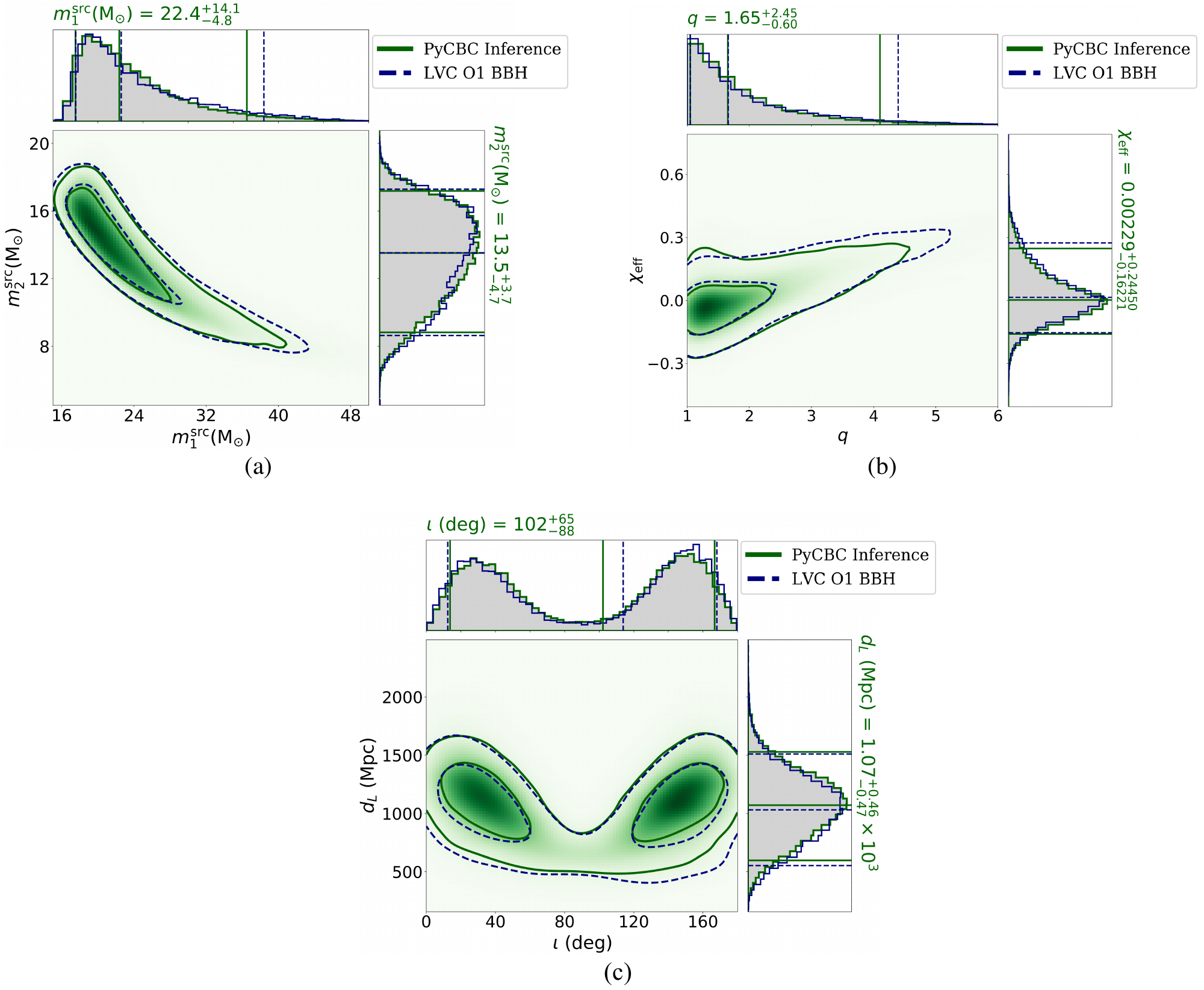}
\caption{\label{fig:lvt151012_plots}Posterior probability densities for the main parameters of interest from the PyCBC Inference analysis of LVT151012. The parameters plotted are (a): $m_1^{\mathrm{src}} - m_2^{\mathrm{src}}$, (b): $q - \chi_{\mathrm{eff}}$ (c): $d_{L} - \iota$.  The bottom-left panel in each of (a), (b) and (c) show two-dimensional probability densities with 50\% and 90\% credible contour regions from the PyCBC Inference posteriors. The top-left and the bottom-right panels in each figure show one-dimensional posterior distributions for the individual parameters with solid lines at the 5\%, 50\% and 95\% percentiles. For comparison, we also show 50\% and 90\% credible regions, one-dimensional posterior probabilities with dashed lines at 5\%, 50\% and 95\% percentiles using the posterior samples obtained from the LIGO Open Science Center~\cite{Vallisneri:2014vxa} for the LVT151012 analysis reported in Ref.~\cite{TheLIGOScientific:2016pea} using the IMRPhenomPv2 model. The measurements again show that the spins, inclination, and distance are not very well constrained and there is support for the system being both ``face-on'' and ``face-off''. Our measurements are in agreement with the results presented in \cite{TheLIGOScientific:2016pea} within the statistical errors of measurement of the parameters.}
\end{figure*}

Recorded in Table~\ref{tab:results}, is a summary of the median and 90\% credible interval values calculated for GW150914, GW151226, and LVT151012 analyses.
Results for $m_1^{\mathrm{src}} - m_2^{\mathrm{src}}$, $q - \chi_{\mathrm{eff}}$, and $d_L - \iota$ are shown in  Figs.~\ref{fig:gw150914_plots}, \ref{fig:gw151226_plots}, and \ref{fig:lvt151012_plots} for GW150914, GW151226, and LVT151012 respectively.
The two-dimensional plots in these figures show the 50\% and 90\% credible regions, and the one-dimensional marginal distributions show the median and 90\% credible intervals.
Overlaid are the one-dimensional marginal distributions, median, and 90\% credible intervals, as well as the 50\% and 90\% credible regions using the samples obtained from the LIGO Open Science Center~\cite{Vallisneri:2014vxa} for the analyses of the three events reported in Ref.~\cite{TheLIGOScientific:2016pea} using the IMRPhenomPv2 model for comparison.
The results show that GW150914 has the highest mass components among the three events. GW150914 has more support for equal mass ratios whereas the GW151226 and LVT151012 posteriors support more asymmetric mass ratios. Overall, there is preference for smaller spins, with GW151226 having the highest spins among the three events. While the inclination and luminosity distances are not very well constrained, with generally a support for both ``face-on'' ($\iota = 0$, line of sight parallel to binary angular momentum) and ``face-off'' ($\iota = \pi$, line of sight anti-parallel to binary angular momentum) systems for all the three events, and GW150914 seem to have more preference for face-off systems.
We also computed $\chi_{p}$ for each of the three events and found no significant measurements of precession.
Overall, our results are in agreement with those presented in Ref.~\cite{TheLIGOScientific:2016pea} within the statistical errors of measurement.

\begin{table*}
\centering\begin{tabular}{lccc} 
\hline
Parameter & GW150914 & GW151226 & LVT151012 \\
\hline\hline
$\mathcal{M}^{\mathrm{det}}$ & $31.0^{+1.6}_{-1.5}$~M$_{\odot}$ & $9.7^{+0.06}_{-0.06}$~M$_{\odot}$ & $18.1^{+1.0}_{-0.7}$~M$_{\odot}$ \\
$m_1^{\mathrm{det}}$ & $38.8^{+5.4}_{-3.3}$~M$_{\odot}$ & $15.0^{+8.4}_{-3.4}$~M$_{\odot}$ & $27.0^{+16.5}_{-5.6}$~M$_{\odot}$ \\
$m_2^{\mathrm{det}}$ & $32.9^{+3.2}_{-4.9}$~M$_{\odot}$ & $8.4^{+2.3}_{-2.6}$~M$_{\odot}$ & $16.3^{+4.2}_{-5.8}$~M$_{\odot}$ \\
$\mathcal{M}^{\mathrm{src}}$ & $28.2^{+1.6}_{-1.4}$~M$_{\odot}$ & $8.9^{+0.3}_{-0.25}$~M$_{\odot}$ & $15.0^{+1.3}_{-1.0}$~M$_{\odot}$ \\
$m_1^{\mathrm{src}}$ & $35.3^{+5.0}_{-3.1}$~M$_{\odot}$ & $13.7^{+7.7}_{-3.2}$~M$_{\odot}$ & $22.4^{+14.1}_{-4.8}$~M$_{\odot}$ \\
$m_2^{\mathrm{src}}$ & $29.9^{+3.0}_{-4.4}$~M$_{\odot}$ & $7.7^{+2.1}_{-2.4}$~M$_{\odot}$ & $13.5^{+3.7}_{-4.7}$~M$_{\odot}$ \\
$q$ & $1.17^{+0.38}_{-0.16}$ & $1.78^{+2.21}_{-0.71}$ & $1.65^{+2.45}_{-0.6}$ \\
$\chi_{\mathrm{eff}}$ & $-0.033^{+0.11}_{-0.12}$ & $0.2^{+0.18}_{-0.07}$ & $0.0023^{+0.24}_{-0.16}$ \\
$a_1$ & $0.29^{+0.57}_{-0.26}$ & $0.53^{+0.37}_{-0.45}$ & $0.28^{+0.51}_{-0.26}$ \\
$a_2$ & $0.33^{+0.56}_{-0.30}$ & $0.51^{+0.43}_{-0.46}$ & $0.40^{+0.51}_{-0.36}$ \\
$d_L$ & $497^{+126}_{-202}$~Mpc & $454^{+164}_{-187}$~Mpc & $1071^{+458}_{-473}$~Mpc \\
\hline
\end{tabular}
\caption{Results from PyCBC Inference analysis of GW150914, GW151226, and LVT151012. Quoted are the median and 90\% credible interval values for the parameters of interest. Interpretations of these results are summarized in Sec.~\ref{sec:astro_events}.}
\label{tab:results}
\end{table*}

\section{Conclusions}\label{sec:con}

In this paper we have described PyCBC Inference, a Python-based toolkit with a simplified interface for parameter estimation studies of compact-object binary  mergers.
We have used this toolkit to estimate the parameters of the gravitational-wave events GW150914, GW151226, and LVT151012; our results are consistent with previously published values. In these analyses, we do not marginalize over calibration uncertainty of the measured strain in our results, which was included in prior work, for example Refs.~\cite{TheLIGOScientific:2016pea, TheLIGOScientific:2016wfe,Abbott:2016izl}. We will implement this in PyCBC Inference in the future.
We have made the samples of the posterior probability density function from the PyCBC Inference analysis of all three events available in Ref.~\cite{ResultsRepo} along with the instructions and configuration files needed to replicate these results.
The source code and documentation for PyCBC Inference is available as part of the PyCBC software package at \url{http://pycbc.org}. 

PyCBC Inference has already been used to produce several astrophysical results: (i) a test of the black hole area increase law~\cite{PhysRevD.97.124069}, (ii) measuring the viewing angle of GW170817 with electromagnetic and gravitational-wave signals~\cite{Finstad:2018wid}, and (iii) measuring the tidal deformabilities and radii of neutron stars from the observation of GW170817~\cite{De:2018uhw}.
The results presented in this paper and in the studies above demonstrate the capability of PyCBC Inference to perform gravitational-wave parameter estimation analyses.
Future developments under consideration are implementation of models to marginalize over calibration errors, generic algorithms to perform model selection, HPD to compute credible intervals, and methods for faster computation of the likelihood.

\section{Acknowledgements} \label{sec:ack}
The authors would like to thank Will Farr and Ben Farr for valuable insights into the intricacies of ensemble MCMCs. We also thank Ian Harry, Christopher Berry, and Daniel Wysocki for helpful comments on the manuscript. This work was supported by NSF awards PHY-1404395 (DAB, CMB), PHY-1707954 (DAB, SD), and PHY-1607169 (SD). Computations were supported by Syracuse University and NSF award OAC-1541396. We also acknowledge the Max Planck Gesellschaft for support and the Atlas cluster computing team at AEI Hannover. DAB thanks the \'Ecole de Physique des Houches for hospitality during the completion of this manuscript. The authors thank the LIGO Scientific Collaboration for access to the data and acknowledge the support of the United States National Science Foundation (NSF) for the construction and operation of the LIGO Laboratory and Advanced LIGO as well as the Science and Technology Facilities Council (STFC) of the United Kingdom, and the Max-Planck-Society (MPS) for support of the construction of Advanced LIGO. Additional support for Advanced LIGO was provided by the Australian Research Council. This research has made use of data obtained from the LIGO Open Science Center \url{https://losc.ligo.org}.

\section*{References}

\providecommand{\newblock}{}

\end{document}